\documentstyle[12pt]{article}
\begin{document}

\noindent
{\Large\bf On a Three-Dimensional Gravity Model with Higher Derivatives}
\\[3mm]

\noindent
{\bf Carlos Pinheiro$^{1}$, Gentil O. Pires$^{2}$ and Claudio Sasaki$^{3}$}\\

\ \\[2mm]

\hspace*{3cm}\hfill \
\begin{minipage}{9.0cm}
\rule{9.0cm}{0.5mm}
{\footnotesize The purpose of this work is to present a model for 3D 
massive gravity with topological and higher-derivative terms. 
Causality and unitarity are discussed at tree-level.
Power-counting renormalizability is also contemplated.}
\rule{9.0cm}{0.5mm}
\end{minipage}

\vspace{2cm}

The formulation of a quantum gauge theory for gravity is certainly one of the 
hardest tasks theoretical physics has been facing over the past years. 
The main
difficulty lies in the apparent clash existing between Quantum Mechanics and
General Relativity: the ideas and concepts of Quantum Mechanics, that work consistently
whenever applied to electromagnetism, weak interactions and strong nuclear forces,
lack  consistency in the context of a theory based on General Relativity, at
least in 4 space-time  dimensions, where infinities show up and can not be 
systematically eliminated by renormalization.

A question arises whether gravitation could be correctly
formulated only in terms of a geometrical approach. In other  words, could the
gravitational

\vspace{1cm}

\noindent
\rule{7cm}{0.2mm}\\
\footnotesize{$^{1}$Universidade Federal do Esp\'{\i}rito Santo
(UFES), 
Departamento de F\'{\i}sica, 
 \\ Av. Fernando Ferrari, s/n 29060-900, Campus das
Goiabeiras, \\ Vit\'oria, ES, Brazil.
E-mail: fcpnunes@cce.ufes.br \\

\noindent
$^{2}$Instituto de
F\'{\i}sica, 
Universidade Federal do Rio de Janeiro (UFRJ), Caixa Postal 68528,
21945-970, Rio de Janeiro, RJ, Brazil. E-mail: gentil@if.ufrj.br \\

\noindent
$^{3}$Centro Brasileiro de Pesquisas F\'{\i}sicas (CBPF/CNPq), Rua
Dr. Xavier Sigaud, 150, 22290-180, Rio de Janeiro, RJ, Brazil.} 

\newpage
\normalsize
\noindent
field be described by gravitons,  instead of a metric
that carries the information on the dynamics of gravitation? There is a school of thought,  
represented by R. Penrose, that conjectures that Quantum Mechanics should be modified so as to
encompass General Relativity, in such a way that the formulation of a quantum
theory for gravitation is made possible in 4 dimensions.

Nevertheless, rather than adopting here Penrose's viewpoint, we shall proceed
along different lines. We consider the usual Einstein-Hilbert action for gravitation
modified by the introduction of extra terms that work to render a
theory for gravitation  quantum-mechanically 
consistent.  In this framework, one chooses  the metric fluctuations around flat-space metric as the quantum
field.

Following this approach, systematic studies have been carried out over the past
decade. Linearized Einstein gravity theory in D dimensions, 
Einstein gravity with Chern-Simons
term, gravity with a Proca-like mass, or a cosmological constant and a mixed
Einstein-Chern-Simons-Proca model, are some examples of gravitational models
proposed in the gauge field approach.

The basic requirements in order to select or rule out  a quantum model for gravity
are unitary and renormalizability. The models mentioned above \cite{1,2,3} have in
common the fact that all Lagrangians are at most of second order in space-time
derivatives.

Our main task here is to treat a higher-derivative model for gravitation in the 
presence of a Chern-Simons topological term, which then selects 3 dimensions.
The questions we are left with concern the study of unitary and power-counting
renormalizability for such models. Also, we already know that in 4 dimensions
the introduction of higher derivatives does not suffice to reconcile unitarity 
and renormalizability for gravity models \cite{4}.

Therefore, motivated to better understand the mismatch between unitarity and
renormalizability in quantum gravity, we shall try to formulate a 
power-counting
renormalizable, and possibly unitary, quantum field theory for 3D-gravity in the presence of a topological Chern-Simons mass term.

Let us then start by proposing a gravity theory described by the following Lagrangian density:

\begin{equation}
L = L_{H.E} + L_{\lambda} + L_{C.S}
\end{equation}
where the terms in the r.h.s. are respectively the usual Einstein-Hilbert term, the
quartic-derivative Lagrangian and the usual Chern-Simons topological contribution:

\begin{eqnarray}
&& L_{H.E} = -\frac{1}{2\kappa^2} \sqrt{-g} R , \\ 
&& L_{\lambda} = \frac{\lambda}{2} \sqrt{-g} R^2 , \\
&& L_{C.S} = \frac{1}{2\mu} \varepsilon^{\mu\nu\lambda} \left(R^{\alpha}_{\beta\mu\nu}
\Gamma^{\beta}_{\alpha\lambda} - \frac{2}{3} \Gamma^{\alpha}_{\beta\mu}
\Gamma^{\beta}_{\gamma\nu}
\Gamma^{\gamma}_{\alpha\lambda}\right) , 
\end{eqnarray}
where $R^{\alpha}_{\beta\mu\nu}$ stands for the Riemann tensor,
$\Gamma^{\alpha}_{\beta\mu}$ represents the Christoffel symbols 
and $R$ denotes the curvature scalar.

The gauge-fixing functional here corresponds to the De Donder gauge choice,

\begin{equation}
L_{g.f} = -\frac{1}{2\alpha} F_{\mu}F^{\mu},
\end{equation} 
where $\alpha$ is the gauge-fixing parameter and

\begin{equation}
F_{\mu} [h_{\rho\sigma}] = \partial_{\lambda} \left(h^{\lambda}_{\mu} -
\frac{1}{2} \delta^{\lambda}_{\mu} h^{\nu}_{\nu}\right)\ .
\end{equation}

One expands the metric tensor around the flat background and the linear 
fluctuation is taken as the quantum field,

\begin{equation}
g_{\mu\nu} (x) = \eta_{\mu\nu} + \kappa h_{\mu\nu} (x) \ .
\end{equation}
Here, $k$ is the Newton gravitational constant.

Once the field parametrization (7) has been plugged into (1), the following
bilinear Lagrangian comes out:

\begin{equation}
L = \frac{1}{2} h^{\mu\nu}\theta_{\mu\nu ,\kappa\lambda} h^{\kappa\lambda} ,
\end{equation}
where the operator $\theta_{\mu\nu ,\kappa\lambda}$ is 
expressed in terms of the 
extended spin operators \cite{2} according to

\begin{eqnarray}
\theta_{\mu\nu ,\kappa\lambda} & =  &\frac{\Box}{2} P^{(0)}
+ \frac{\Box}{2\alpha} P^{(1)}_m - 
\left[\left(\frac{4\alpha -3}{4\alpha}\right)
\Box - 3 \lambda \kappa^2\Box^2\right]P^{(0)}_s + \nonumber \\
& + &\frac{\Box}{4\alpha} P^{(0)}_w - \frac{\sqrt{3}}{4\alpha} \Box
 P^{(0)}_{sw} 
-\frac{\sqrt{3}}{4\alpha} \Box P^{(0)}_{ws} + \frac{1}{2M} (S_1+S_2) ,
\end{eqnarray}
where we have introduced the parameter $M=\frac{\mu}{8\kappa^2}$.

By adopting the Feynman gauge $(\alpha =1)$, (9) can be rewritten as

\begin{eqnarray}
\theta_{\mu\nu ,\kappa\lambda} & = & \frac{\Box}{2} P^{(2)} +
\frac{\Box}{2} P^{(1)}_m - \left(\frac{\Box}{4} - 3\lambda\kappa^2\Box^2\right)
P^{(0)}_s + \nonumber \\
& + & \frac{\Box}{4} P^{(0)}_w 
- \frac{\sqrt{3}}{4} \Box P^{(0)}_{sw} - 
\frac{\sqrt{3}}{4} \Box P^{(0)}_{ws} + \frac{1}{2M}
(S_1+S_2)  .
\end{eqnarray}
Notice that in the case $\lambda = 0$ and $M\rightarrow \infty$, the operator
$\theta$ corresponding to the usual Einstein-Hilbert case is recovered \cite{2}.

The corresponding propagator can be obtained from the following generating
functional:

\begin{equation}
W[T_{\rho\sigma}] = -\frac{1}{2} \int d^3 x d^3 y T^{\mu\nu} \theta^{-1}_{\mu\nu ,\kappa\lambda} 
T^{\kappa\lambda} ,
\end{equation}
where $(\theta^{-1})_{\mu\nu ,\kappa\lambda}$ is the inverse operator of (10) and the  propagator reads

\begin{equation}
\langle T[h_{\mu\nu}(x) h_{\kappa\lambda} (y)]\rangle =
i\theta^{-1}_{\mu\nu ,\kappa\lambda} \delta^3 (x-y) .
\end{equation}
With the help of the multiplicative table for the extended Barnes-Rivers operators
\cite{2,4}, $\theta^{-1}$ is found to be given by

\begin{eqnarray}
&&\hspace{3cm}\theta^{-1}_{\mu\nu ,\kappa\lambda} = X_1 P^{(2)} +
X_2P^{(1)}_m + X_3P^{(0)}_s + X_4P^{(0)}w + \nonumber \\
&&\hspace{3cm}+X_5P^{(0)}_{sw} + X_6P^{(0)}_{ws} +
X_7S_1 + X_8S_2 \ . \hspace{2.3cm} \mbox{(13.a)} \nonumber
\end{eqnarray}

\noindent
The coefficients $X_1\cdots X_8$ are found to have the following expressions:
\begin{eqnarray}
&&\hspace{4.5cm}X_1 = \frac{2M^2}{\Box (M^2+\Box)}\ ,  \nonumber
\hspace{3.6cm} \mbox{(13.b)} \\
&& \hspace{4.5cm}X_2 = \frac{2}{\Box}\ , \nonumber \hspace{5.4cm}  \mbox{(13.c)}\\
&&\hspace{2.3cm} X_3 = \frac{-[(4M^2\lambda\kappa^2 - 3)\Box - 4M^2]}{\Box [4\lambda\kappa^2 \Box^2 +
(4M^2\lambda\kappa^2-1) \Box - M^2]}\ , \nonumber \hspace{1.5cm} 
\mbox{(13.d)}\\
&& \hspace{2.3cm}X_4 = \frac{16\lambda\kappa^2}{4\lambda\kappa^2\Box-1}\ , \nonumber
\hspace{6.0cm}\mbox{(13.e)} \\
&&\hspace{2.3cm} X_5 = \frac{2\sqrt{3}}{\Box (4\Box \lambda\kappa^2 - 1)}\ , \nonumber
\hspace{5.3cm} \mbox{(13.f)} \\
&& \hspace{2.3cm}X_6 = \frac{2\sqrt{3}}{\Box (4\Box \lambda\kappa^2 - 1)}\ , \nonumber
\hspace{5.3cm} \mbox{(13.g)}\\
&&\hspace{2.3cm} X_7 =  \frac{- 2M}{\Box^2 (M^2+\Box )}\ , \nonumber \hspace{5.5cm} \mbox{(13.h)}\\
&&\hspace{2.3cm} X_8 =  \frac{- 2M}{\Box^2 (M^2+\Box )}\ . \nonumber \hspace{5.6cm} \mbox{(13.i)}
\end{eqnarray}
By use of (13), the propagator for the $h_{\mu\nu}$ field is finally known.
In the case $\lambda =0$, the Einstein-Chern-Simons theory is recovered as in
\cite{2}, since the model presented here also displays gauge
invariance. If, as well as 
$\lambda =0$, the limit $M\rightarrow \infty$ is taken, the pure Einstein gravity    
in 3D is reobtained. Now, if only the limit $M\rightarrow \infty$ is considered, the propagator
(12) describes the higher derivative gravitational theory. This is so because
no mass term is present in (1) that explicitly breaks gauge symmetry.

We will now  discuss the tree-level unitarity of the model. To do that,
we couple the propagator (12) to external currents, $T_{\mu\nu}$, 
compatible with
the symmetry of the theory and then we analyze the imaginary part of 
the residue of the current amplitude at the poles.

Let this amplitude be
\setcounter{equation}{13}
\begin{equation}
A = T^{\mu\nu *}(k) \langle T[h_{\mu\nu}(-k)h_{\kappa\lambda}
(k)]\rangle T^{\kappa\lambda} \ .
\end{equation}

By inserting the propagator (13) in eq. (14), and taking the  
 residues at the poles, one can check that only the spin projectors
$P^{(2)}$ and $P^{(0)}_s$ survive, by virtue of current
conservation:

\begin{equation}
\omega T = 0.
\end{equation}

Henceforward, the procedure is analogous to that in \cite{2}, 
and we analyze the poles in the spin-2 and spin-0 sectors. 
Let us find these poles.

Going over to momentum space, one gets

\begin{equation}
X_1 = \frac{2M^2}{k^2(k^2-M^2)}\  .
\end{equation}
This gives two poles for the spin-2 sector,

\begin{equation}
k^2 = 0 \quad , \quad k^2 = M^2 \ . 
\end{equation}

Analogously, for the spin-0 sector, one has

\begin{equation}
X_3 = \frac{-[(4M^2\lambda\kappa^2-3)k^2+4M^2]}{4\lambda \kappa^2(k^2)^3-
(4M^2\lambda\kappa^2-1)(k^2)^2-M^2k^2}\ ,
\end{equation}
whose poles are found to be given by the roots of the cubic equation

\begin{equation}
4\lambda\kappa^2 (k^2)^3 - (4M^2\lambda\kappa^2-1) (k^2)^2 -M^2k^2 = 0 \ .
\end{equation}
We then find

\begin{equation}
k^2 = 0 \ , \quad \quad k^2 = M^2 \ ,  \quad
k^2 = \frac{- 1}{4\lambda \kappa^2} \ .
\end{equation}

\noindent
Therefore, the theory leads to a massless and two massive excitations. One has
to take $\lambda <0$ in order to avoid a tachyon in the spectrum.

By considering (14), and taking into account that the analysis is performed
at the poles in $k^2$, it can be checked that

\begin{eqnarray}
&&\hspace{3cm} Im \ Res \ A|_{k^2=0} = 0 \ , \hspace{4.9cm}
\mbox{(21.a)}\nonumber \\
&&\hspace{3cm} Im \ Res \ A|_{k^2=M^2} > 0 \ , \hspace{4.6cm}
\mbox{(21.b)}\nonumber\\
&&\hspace{3cm} Im \ Res |_{k^2 = \frac{-1}{4\lambda\kappa^2}} > 0 \ .\hspace{4.9cm} \mbox{(21.c)}\nonumber
\end{eqnarray}

This suggest that the massless excitation is not a dynamical degree 
of freedom;
on the other hand, along with this non-dynamical mode, there appear two physically
acceptable massive quanta that propagate in 3 space-time dimensions.
So, since both are gauge-invariant, a theory
with the action (1) behaves in the same way as  3D Einstein-Chern-Simons model,
i.e., only massive poles do propagate.  From (21), one
concludes that the theory does not carry negative-norm states, which is a necessary condition
for unitarity. On the other hand, from (16) and (18), one gets that the higher-derivative
model is renormalizable in 3D, the asymptotic behavior of its propagator being
of the type $\frac{1}{k^4}$. Ghosts and tachyons are absent, since $\lambda <0$ automatically leads
to the condition $Im \ Res \ A>0$ together with non-negative definite
poles in $k^2$. It is worth mentioning that, although the term
$[k^2(k^2-M^2)]^{-1}$ appears in the graviton propagator, here in 3D (contrary
to what happens in 4D) a ghost does not show up, in view of  eq. (21.a): the
massless pole does not describe any propagating degree of freedom.

We can enrich the model by adding an independent higher-derivative
term of the form

\setcounter{equation}{21}

\begin{equation}
L_{\xi} = \frac{\xi}{2} \sqrt{-g} R_{\mu\nu} R^{\mu\nu} \ ,
\end{equation}
to be adjoined to (1) with the gauge-fixing of eqs. (5) and (6).

Equations (8) can now be suitably rewritten as

\begin{equation}
L = \frac{1}{2} h^{\mu\nu}\theta_{\mu\nu ,\kappa\lambda} h^{\kappa\lambda} \ , 
\end{equation}
with the operator $\theta_{\mu\nu ,\kappa\lambda}$ being given  by

\begin{eqnarray}
&& \theta_{\mu\nu ,\kappa\lambda} = \Box \left(\frac{1}{2} +
\frac{\xi \kappa^2}{4}\Box \right) P^{(2)} +
\frac{1}{2\alpha} \Box P^{(1)}_m + \Box (-1 + \frac{3}{4\alpha} +
\xi\kappa^2 \Box + \nonumber \\
&& +3\lambda\kappa^2 \Box )
P^{(0)}_s 
+ \frac{1}{4\alpha}\Box P^{(0)}_w  
- \frac{\sqrt{3}}{4\alpha} \Box P^{(0)}_{sw}
- \frac{\sqrt{3}}{4\alpha}\Box P^{(0)}_{ws} +
\frac{1}{2M} (S_1+S_2) \ . \nonumber \\
&& ~~ 
\end{eqnarray}

Again, if $\xi$ and $\lambda$ are both vanishing, one 
recovers the operator for pure
Einstein gravity in $D=3$. The term in (22) is gauge invariant, so that
if $\xi =0$, we reobtain the model described by (1).  

Following the same steps as already illustrated previously, the inverse operator
can be found with the help of the multiplicative table as in \cite{2,4}. It is
given by an expression of the form of eq. (13), with the difference that the
coefficients $X_1\cdots X_8$ now read

\begin{eqnarray}
&& \hspace{1.6cm} X_1 = \frac{4M^2(\xi\kappa^2\Box + 2)}{[M^2\xi^2\kappa^4\Box^2 + 
(4M^2\xi\kappa^2+4)\Box +4M^2)\Box}\ , \hspace{1.3cm} \mbox{(25.a)} \nonumber  \\
&& \hspace{1.6cm}X_2 = \frac{2}{\Box}\ ,  \hspace{8.0cm} \mbox{(25.b)} \nonumber\\
&&\hspace{1.6cm} X_3 = \frac{B}{C}\ , \hspace{8.0cm} \mbox{(25.c)} \nonumber
\end{eqnarray}

\noindent where
$ B = (-8)[2M^2\lambda\xi\kappa^4\Box^2 + 
    (-2M^2\xi\kappa^2 - 3 + 4M^2\lambda\kappa^2) \Box - 4M^2] 
$

\noindent and  
$ C = \Box [(8M^2\xi^2\kappa^6\lambda + 3M^2\xi^3\kappa^6)\Box^3 +
(12\xi\kappa^2 + 10M^2\xi^2\kappa^4  
+ 32\lambda\kappa^2 + + 32M^2\lambda\xi\kappa^4 )\Box^2 
+ (-8 + 32M^2 \lambda\kappa^2 + 4M^2\xi\kappa^2)\Box - 8M^2] \ , 
$

\begin{eqnarray}
&& \hspace{1.6cm}X_4 = \frac{4(3\xi
\kappa^2+8\lambda\kappa^2)}{[(3\xi\kappa^2+8\lambda\kappa^2)\Box-2]}\,
\hspace{4.7cm} \mbox{(25.d)} \nonumber \\
&&\hspace{1.6cm} X_5 = \frac{4\sqrt{3}}{\Box
[(3\xi\kappa^2+8\lambda\kappa^2)\Box - 2]}\ ,
\hspace{4.4cm} \mbox{(25.e)} \nonumber  \\
&&\hspace{1.6cm} X_6 =   \frac{4\sqrt{3}}{\Box
[(3\xi\kappa^2+8\lambda\kappa^2)\Box - 2]}\ , 
\hspace{4.4cm} \mbox{(25.f)} \nonumber \\
&&\hspace{1.6cm} X_7 =  \frac{- 8M}{\Box^2[M^2\xi^2\kappa^4\Box^2 + (4M^2\xi\kappa^2+4) \Box +
4M^2]}\ ,  \hspace{1.3cm} \mbox{(25.g)} \nonumber  \\
&&\hspace{1.6cm}  X_8  = \frac{- 8M}{\Box^2[M^2\xi^2\kappa^4\Box^2 + (4M^2\xi\kappa^2+4) \Box +
4M^2]} \ .\hspace{1.3cm} \mbox{(25.h)} \nonumber
\end{eqnarray}

One can immediately check that, if $\xi =\lambda = 0$ 
and $M\rightarrow \infty$, the
propagator for the pure Einstein theory in $D=3$ can be recovered in  the Feynman gauge 
$(\alpha = 1)$.

The discussion of unitarity follows the same steps as in the previous case,
with the difference that the coefficients  $X_1$ and $X_3$ are 
more cumbersome, which
renders the algebraic derivation of the poles more involved.

However, as already expected due to gauge invariance, a massless pole
is present in the spin-2
sector; the massive poles are given as the roots of a quadratic
equation in 
$k^2$. As for the spin-0 sector, it also displays a massless pole along with
massive poles that appear as the roots of a cubic equation.

The full theory is gauge-invariant and, due to the fact that gravitational
effects in $D=3$ are global ones, the massless pole corresponds to a non-propagating  
degree of freedom. On the other hand, the massive gravitons propagate 
as in pure Einstein-Chern-Simons 
model: negative-norm states do not appear that spoil the spectrum, 
which does 
not affect unitarity.

The asymptotic behavior of the propagator is also of the form $k^{-4}$ and, based on
the results of \cite{1}, one can guarantee the renormalizability of the theory
in an arbitrary gauge. In $D=3$, the clash between unitarity and renormalizability does not
show up. In $D=4$, by setting conditions on the parameters $\xi$ and $\lambda$ so as to  ensure
unitarity, renormalizability is unavoidably lost. If one sticks to the latter, unitarity
is then lost \cite{4}. In $D=3$, the behavior of the models presented here is actually
very close to the main aspects of the Eisntein-Chern-Simons theory \cite{2}.

\section*{Acknowledgements}

The authors thank Dr. J.A. Helay\"{e}l Neto for discussions, 
suggestions and careful 
reading of the manuscript. The authors are partially supported by the 
Conselho Nacional de Desenvolvimento Cient\'{\i}fico e  Tecnol\'ogico, CNPq-Brazil.

\end{document}